\documentclass[letterpaper, 10 pt, conference]{ieeeconf}  % Comment this line out if you need a4paper

\IEEEoverridecommandlockouts                              % This command is only needed if 
\usepackage{graphicx} % for pdf, bitmapped graphics files
\usepackage{amsmath} % assumes amsmath package installed

\title{\LARGE \bf
Crackle Detection In Lung Sounds Using Transfer Learning And Multi-Input Convolitional Neural Networks
}

\author{Truc Nguyen and Franz Pernkopf% <-this % stops a space
\thanks{*This work was supported by the Vietnamese - Austrian Government scholarship and the Austrian Science Fund (FWF) under the 
project number I2706-N31.}% <-this % stops a space
\thanks{Truc Nguyen and Franz Pernkopf are with Signal Processing and Speech Communication Lab., Graz University of Technology, Austria
        {\tt\small t.k.nguyen@tugraz.at, pernkopf@tugraz.at}}%
}

\begin{document}

\maketitle
\thispagestyle{empty}
\pagestyle{empty}

%%%%%%%%%%%%%%%%%%%%%%%%%%%%%%%%%%%%%%%%%%%%%%%%%%%%%%%%%%%%%%%%%%%%
\begin{abstract}

Large annotated lung sound databases are publicly available and might be used to train algorithms for diagnosis systems. However, it might be a challenge to develop a well-performing algorithm for small non-public data, which have only a few subjects and show differences in recording devices and setup. In this paper, we use transfer learning to tackle the mismatch of the recording setup. This allows us to transfer knowledge from one dataset to another dataset for crackle detection in lung sounds. In particular, a single input convolutional neural network (CNN) model is pre-trained on a source domain using ICBHI 2017, the largest publicly available database of lung sounds. We use log-mel spectrogram features of respiratory cycles of lung sounds. The pre-trained network is used to build a multi-input CNN model, which shares the same network architecture for respiratory cycles and their corresponding respiratory phases. The multi-input model is then fine-tuned on the target domain of our self-collected lung sound database for classifying crackles and normal lung sounds. Our experimental results show significant performance improvements of 9.84\% (absolute) in F-score on the target domain using the multi-input CNN model based on transfer learning for crackle detection in adventitious lung sound classification task.
\newline

\indent \textit{Clinical relevance}— Crackle detection in lung sounds, multi-input convolutional neural networks, transfer learning.
\end{abstract}

%%%%%%%%%%%%%%%%%%%%%%%%%%%%%%%%%%%%%%%%%%%%%%%%%%%%%%%%%%%%%%%%%%%%%%%%%%%%%%%%
\section{INTRODUCTION}

Lung sounds are relevant indicators of respiratory health~\cite{sarkar2015auscultation},~\cite{rocha2018alpha}. They are classified as normal and adventitious. Normal respiratory sounds are heard when no respiratory disorders exist. Adventitious lung sounds usually present  pulmonary disorders that are superimposed on the normal respiratory sounds. Crackles are adventitious lung sound, which are discontinuous, explosive and non-musical. They can be fine or coarse depending on their duration, loudness, pitch, timing in the respiratory cycle (i.e. inspiration or expiration). The appearance of crackles may be an early sign of respiratory diseases. The number of crackles per breath is associated with the severity of diseases in patients with interstitial lung conditions. Moreover, the waveform and occurence of crackles may have clinical significance in differential diagnosis of cardiorespiratory conditions~\cite{rocha2018alpha}. 

Auscultation is an important mean to diagnose pulmonary diseases using a stethoscope. In the last decades, computational methods i.e. Computational lung sound analysis (CLSA) have been developed for automated detection and classification of adventitious lung sounds, which use digital recording, signal processing techniques and machine learning algorithms. They potentially overcome the conventional method's limitations and offer advantages for medical diagnosis%~\cite{gurung2011computerized},
~\cite{pramono2017automatic}. Furthermore, they are carefully evaluated in real-life scenarios and can be used as portable easy-to-use devices without the necessity of expert interaction; especially, beneficial when facing infectious diseases as COVID-19. 

In CLSA systems, adventitious lung sound detection and classification includes three key steps: pre-processing of audio signals, extracting the relevant features and detection or classification of adventitious sounds (i.e. crackles, wheezes and both of them).
In the pre-processing step, resampling and filtering are applied to remove heart sounds, background noises and contact interference. Subsequently, features in the time domain and time-frequency domain are extracted such as the Hilbert-Huang transform, Fourier transform, short-time Fourier transform, wavelet transform, and S-transform~\cite{chen2019triple}. The features are processed by conventional machine learning such as self-organizing maps, %~\cite{malmberg1996SOM}, 
Gaussian mixture models (GMMs), %~\cite{bahoura2009patterngmm}, 
support vector machines (SVMs)~\cite{bokov2016wheezingSVM}, 
and multi perceptron networks (MPNs)~\cite{liu2017lung}. 
Recently, convolutional neural network (CNNs)~\cite{demir2020convolutional}, %~\cite{dBardou2018}~\cite{aykanat2017cnn},
~\cite{tNguyen2020embc},~~\cite{pham2020embc}, recurrent neural networks (RNNs)~\cite{messner2018crackle},~\cite{dPerna2019lstm} %~\cite{kochetov2018noise},
or hybrid CNNs and RNNs~\cite{acharya2020deep},~\cite{shi2019lung},~\cite{messner2020multi} using time-frequency representations such as MFCCs and spectrograms have been the most successful approaches. Due to limitations in the amount of available data, the performance and generalization ability of the lung sound classification system may suffer. To deal with these disadvantages, data augmentation and transfer learning from Imagenet~\cite{demir2020convolutional},~\cite{acharya2020deep}, or audio scene datasets~\cite{shi2019lung} have been explored.

In this work, we improve the generalization ability and performance for crackle detection using our multi-channel lung sound database. We propose a new transfer learning approach for a multi-input convolutional neural network. We use the ICBHI 2017 scientific challenge respiratory sound database~\cite{rocha2018alpha} for pre-training the model using log-mel spectrogram features of full respiratory cycles as input. The pre-trained model is then used to build a multiple input CNN served with features from the respiratory cycles and phases (i.e. inspiration and expiration). Furthermore, to strengthen patterns of adventitious lung sounds, sample padding is applied to fill up the uniform length of the respiratory cycle and phases.
The main contributions of the paper are:
\begin{itemize}
	\item We split respiratory cycles into phases and perform sample padding on both of them to enrich the information of adventitious sounds for the lung sound classification system.
	\item We exploit transfer learning in using the pre-trained single input model to build a multi-input CNN model for lung sound classification.
\end{itemize}

The outline of the paper is as follows: In Section II, we introduce the lung sound databases used as source and target domains. In Section III, we present our system. In Section IV, we present the experimental setup including the evaluation metrics and the experimental results. Finally, we conclude the paper in Section V.

\section{MATERIALS}

\subsection{Source Domain}

The ICBHI 2017 database~\cite{rocha2018alpha} consists of 920 annotated audio samples from 126 subjects. The database includes 6898 different respiratory cycles with 3642 normal cycles, 1864 crackles, 886 wheezes, and 506 cycles consisting of both crackles and wheezes.

\subsection{Target Domain}

\begin{figure}[t]
    \centering
    \includegraphics[width=55mm]{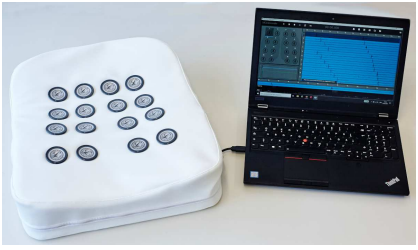}
    \caption{Multi-channel lung sound recording device.}
    \label{fig:recordingdevice}
    \end{figure}

In a clinical trial, the multi-channel lung sound database~\cite{messner2016robust}, \cite{messner2018crackle}, \cite{messner2020multi} has been recorded. It contains lung sounds of 16 healthy subjects and 7 patients diagnosed with idiopathic pulmonary fibrosis (IPF). We used our 16-channel lung sound recording device (see Fig.~\ref{fig:recordingdevice}) to record lung sounds over the posterior chest at two different airflow rates, with 3 - 8 respiratory cycles within 30s. The lung sounds were recorded with a sampling frequency of 16kHz. The sensor signals are filtered with a Bessel high-pass filter with a cut-off frequency of 80Hz and a slope of 24dB/oct. From all recordings, we extracted full respiratory cycles using the airflow signal. Based on the characteristic of fine crackles for IPF from mid to late inspiration~\cite{flietstra2011automated}, we manually annotated respiratory cycles without crackles, as there are several recordings of subjects with IPF, where sensors placed on the top of the multi-channel recording device do not contain crackles. The number of breathing cycles with/without IPF are shown in Table~\ref{number_cycles}.

    \begin{table}[t]
        \renewcommand{\arraystretch}{1.3}
        \caption{Number of subjects and cycles in the dataset}
        \label{number_cycles}
        \centering
        \scalebox{0.8}{
        \begin{tabular}{p{0.2\linewidth}p{0.1\linewidth}p{0.15\linewidth}p{0.15\linewidth}p{0.1\linewidth}} %{c c c c c}
            \hline
            \multicolumn{2}{c}{\textbf{\# Subjects}}  & \multicolumn{3}{c}{\textbf{\# Respiratory Cycles}}\\
            \hline
            \textbf{Healthy} & \textbf{IPF} & \textbf{Normal} & \textbf{Crackles} & \textbf{Total}\\
            
            \hline
            16 & 7 & 4405 & 1791 & 6196 \\
            \hline
        \end{tabular}
    }
    \end{table}

\section{PROPOSED METHODOLOGY}

\begin{figure}[t]
	\centering
	\centerline{\includegraphics[width=60mm]{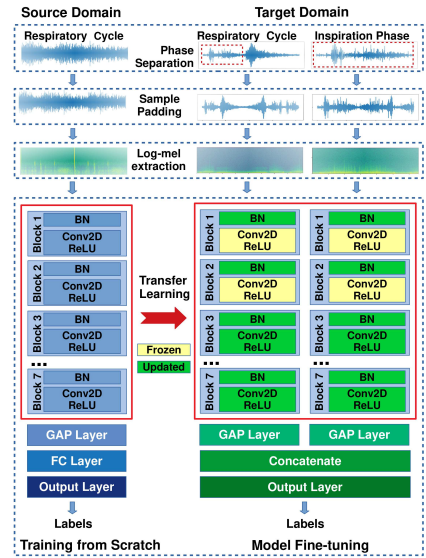}}
	\caption{Proposed System.}
	\label{fig:framework}
\end{figure}

The proposed system includes three key stages shown in Fig.~\ref{fig:framework}. Firstly, the respiratory cycles are pre-processed with separation into respiratory phases and sample padding in the audio signal domain. Secondly, log-mel spectrograms are extracted from the respiratory cycles and phases. Finally, the features are fed to the multi-input CNN model for fine-tuning.

\subsection{Audio Pre-processing and feature extraction}
We use the audio pre-processing and feature extraction techniques of~\cite{tNguyen2020embc} for both source and target domains. As audio recordings from both domains were collected with different sampling rates, respiratory cycles are resampled to 16kHz. Similar to our previous work, the source domain is processed using full respiratory cycles, while the target domain is processed using full respiratory cycles and additionally the respiratory phases (i.e. inspiration/expiration).

\subsubsection{Respiratory Phase Separation}
In addition to monitoring the presence of adventitious sounds, clinicians need to be aware of their timing in the respiratory cycles i.e. early/mid/late inspiratory or expiratory as it may have clinical significance for the assessment of the patient respiratory status and for the differential diagnosis of cardiorespiratory disorders~\cite{jacome2019convolutional}. Therefore, we propose for our CLSA approach to use the combination of a full respiratory cycle with either one or both respiratory phases. To split a respiratory cycle into inspiration and expiration, we use a particular fixed length ratio of inspiration to respiratory cycle. Research about the time duration of respiratory phase for lung cancer patients and dogs~\cite{zhang2014clinical} includes that the average length ratios of inspiration to respiratory cycle is about 1.7(s): 5.2(s), which is approximately 1:3 of inspiration to full cycle. Therefore, we empirically use five length ratios for inspiration to respiratory cycle, namely 1:3, 2:5, 3:7, 4:9, and 1:2. 

\subsubsection{Sample Padding}

The length of the respiratory cycles varies in the data while our CNN model requires the same input sizes. Hence we extract the same number of samples in the respiratory cycle or phase. We choose a fixed length for respiratory cycles by using a maximum length of the respiratory cycles in the target domain (i.e. 131960 samples) and use this length to determine the maximum length of inspiration and expiration using the ratio from above. Furthermore, we partially augment these fixed lengths by sampling from the available cycle or phase samples in time domain. In particular, we do sample padding in time-reversed order to avoid abrupt signal changes%~\cite{tNguyen2020embc}. 
We call this \emph{sample padding}. The sample padding technique showed its efficiency compared to \emph{zero padding} in our previous work~\cite{tNguyen2020embc}. 

Since the target domain consists of IPF subjects' recordings, in which crackles are located in the mid to late inspiratory phase~\cite{sarkar2015auscultation}, we emphasize the crackles (in the mid to late inspiratory phase) by using sample padding in time reversed order, i.e. using the last samples of inspiratory and expiratory phases for padding first.

\subsubsection{Feature Extraction and Normalization}
We use 512 samples as window size of the fast Fourier transform (FFT) without overlap between the windows. The number of mel frequency bins is chosen as 45. The logarithmic scale is applied to the magnitude of the mel spectrograms. The log-mel spectrograms are normalized with zero mean and unit standard deviation. 

\subsection{Data Augmentation}
Similar to our previous work~\cite{tNguyen2020embc}, we use time stretching and  vocal tract length perturbation (VTLP) to balance the training dataset and prevent over-fitting for the source and target domains. 

For the source domain, time stretching is used to double the number of samples of the wheeze, and both crackle and wheeze classes of the training set. VTLP  enlarges the dataset for all classes of the original training set and the time stretched data. More details can be found in~\cite{tNguyen2020embc}. While for the target domain, we use VTLP only for crackles of respiratory cycles and phases of the training set.
\subsection{Deep Learning Approaches}

\subsubsection{Transfer Learning}

In this work, the adventitious lung sound classification tasks for source and target domains are different. The target label space (i.e. normal and crackles) is a subset of the source's label space (i.e. normal, crackles, wheezes and both crackles and wheezes). Furthermore, the energy distributions of the log-mel spectrograms of both domains are different, which is caused by differences in recording devices and noise levels of the audio signal. 
Therefore, transfer learning is promising to deal with the problem of limited training data in our target domain and to enhance the generalization ability and performance of our lung sound classification system. To do so, a CNN model is trained from scratch from the source domain data. Then the pre-trained model is transfered to the target domain by reusing the learned features of the pre-trained model to build a multi-input CNN model. Several of the first layers of the multi-input CNN are frozen while the remaining layers are fine-tuned with the target domain data. We report the performance of transfer learning by fine-tuning at different layers.  

\subsubsection{Pre-trained Convolutional Neural Network (CNN)}
We reuse the CNN model of 7 convolutional compositions in our previous work~\cite{tNguyen2020embc}.
The seven convolutional compositions (i.e. blocks) include a batch normalization layer (BN) and a convolution layer (Conv2D) using ReLU activations (BN-Conv2D-ReLU) shown in the Fig.~\ref{fig:framework}.
The CNN model is trained from scratch using data from the source domain.

\subsubsection{Multi-Input Convolutional Neural Networks (MI-CNNs)}
In order to combine the respiratory cycle and phases of the proposed system, we exploit transfer learning different to   ~\cite{demir2020convolutional},~\cite{acharya2020deep}, or~\cite{shi2019lung}, in which LSTM, or CNN layers were added after the feature learning part of the pre-trained model. In our work, we build MI-CNN models with two or three branches corresponding to the different combination of respiratory phases. There are four combinations such as full cycle - inspiration phase, full cycle - expiration phase, and inspiration - expiration phase and a triplet of full cycle - inspiration - expiration phase. The corresponding feature are fed into each branch of the MI-CNN. Each branch has the same structure reused from the pre-trained model. Following the last convolutional blocks, again global average pooling (GAP) layers are used. The output of each branch is concatenated before fed to the output layer using the softmax activation for classification. The architecture of an MI-CNN model using a pair of full respiratory cycle and inspiration phase is shown in Fig.~\ref{fig:framework}.

Since the data distributions of the source and target domains are different, we update all BN layers of the MI-CNN model.
We freeze the first convolutional blocks and perform fine-tuning of the MI-CNN model beginning with layers of the second, third or fourth convolutional blocks. 

\section{EXPERIMENTS}

\subsection{Setup}
We perform a respiratory cycle-wise classification of crackles and normal classes.
We evaluate the performance for the target domain using Precision $P_+$, Sensitivity $Se$ and F-score~\cite{messner2018crackle}. %(Eqs.~\ref{eqn:evaluation}).
For the target domain, due to the limited amount of data samples, we use 7-fold cross-validation with the recordings of each IPF subject appearing once in the test set. Each subject is assigned to either training, validation or test set. 
The reported performance of the system is an average accuracy of five independent runs for seven folds using the same data splittings.

Training the networks is carried out by optimizing the focal loss using the Adam optimizer at a learning rate of 0.0001 and a batch size of 32 for the source domain and 15 for the target domain. The number of epochs is set to 150 for all tests and the optimal model is that with the highest validation accuracy. We use the Glorot uniform initializer for the network weights. Weight decay regularizer L2 is included at a factor of 0.001. Data is shuffled between the epochs. 

We observe the impact of five separation ratios of inspiration and respiratory cycles. Furthermore, several combinations of respiratory cycle (\textit{\_Cyc\_}), inspiration (\textit{\_Ins\_}) and expiration (\textit{\_Exp\_}) phase for multi-input models have been evaluated. 

\subsection{Performance}

Table~\ref{tbl:comparisonresults} presents the performance comparison of our proposed systems on target domain data using models trained from scratch and via transfer learning using the ICBHI 2017 database for different input combinations using sample padding and zero padding. 
We can see that sample padding (\textit{$\_SamplePad$}) mostly outperforms zero padding (\textit{$\_ZeroPad$}) for single input and the best combinations of multiple inputs and splitting phase ratios. The multi-input CNN systems outperform the scratch single input CNN systems significantly. Transfer learning for multi-input CNN models achieves better performances compared to models trained from scratch. The best performing system (\textit{$2ndConv\_Cyc\_Inp\_Ratio\_12\_SamplePad$}) use transfer learning for the multi-input model of respiratory cycle and inspiration phase with splitting phase ratio of 1:2. Fine-tuning from the second convolutional layer (\textit{$2ndConv\_$}) of sample padding is performed. The F-score is 84.71\%.

\begin{table}[t]
	\renewcommand{\arraystretch}{1.3}
	\caption{Comparison of proposed systems.}
	\label{tbl:comparisonresults}
	\centering
%	\footnotesize
	\scalebox{0.8}{
	\begin{tabular}{l c c l}
		\hline
		\textbf{Proposed Systems} & \textbf{Se} & \textbf{P{+}} & \textbf{F-Score} \\
		\hline
		$Scratch\_Cyc\_SamplePad$ & \textbf{0.8526} &\textbf{0.6675} & \textbf{0.7487} \\
		$Scratch\_Cyc\_Inp\_Ratio\_25\_SamplePad$ & 0.9080 &0.7373 & 0.8137 \\
		$Scratch\_Cyc\_Exp\_Ratio\_49\_SamplePad$ & 0.8975 &0.6926 &  0.7818\\
		$Scratch\_Ins\_Exp\_Ratio\_25\_SamplePad$ & \textbf{0.9035} &\textbf{0.7582} &\textbf{0.8245}  \\
		$Scratch\_Cyc\_Ins\_Exp\_Ratio\_25\_SamplePad$ & 0.8996 &0.7261 & 0.8036 \\
		\hline
		$Scratch\_Cyc\_ZeroPad$ & 0.8450 & 0.6521 & 0.7361 \\
		$Scratch\_Cyc\_Inp\_Ratio\_25\_ZeroPad$ & 0.8908 &0.7476 & 0.8129 \\
		$Scratch\_Cyc\_Exp\_Ratio\_49\_ZeroPad$ & 0.9064 &0.6857 &  0.7807\\
		$Scratch\_Ins\_Exp\_Ratio\_25\_ZeroPad$ & 0.9070 &0.7037 &0.7925  \\
		$Scratch\_Cyc\_Ins\_Exp\_Ratio\_25\_ZeroPad$ &0.9005  &0.7457 & 0.8159 \\
		\hline
		\textbf{$2ndConv\_Cyc\_Inp\_Ratio\_12\_SamplePad$}& \textbf{0.8532} &\textbf{0.8411} & \textbf{0.8471} \\
		$4thBN\_Cyc\_Exp\_Ratio\_13\_SamplePad$ &0.8647  &0.7250 & 0.7887 \\
		$3rdBN\_Ins\_Exp\_Ratio\_49\_SamplePad$ & 0.8627 &0.8005 & 0.8304  \\
		$3rdConv\_Cyc\_Ins\_Exp\_Ratio\_49\_SamplePad$ & 0.8845 &0.7917 & 0.8356 \\

		\hline
	\end{tabular}
}
\end{table}

\section{CONCLUSIONS}
This work introduces transfer learning and multi-input convolutional neural networks using combinations of full respiratory cycle and phases for adventitious lung sound classification. We evaluate the effect of different splitting phase ratios for inspiration phase and respiratory cycle. Furthermore, various combinations of the respiratory cycle and its corresponding phases for the multi-input model are trained from scratch and transfered from the ICBHI 2017 domain to the target domain. Our best transfered system use the multi-input model and performs fine-tuning starting at the second convolutional layer. It uses the respiratory cycle and the inspiration phase with a splitting phase ratio of 1:2. It outperforms the model learned from scratch on the target domain using the full respiratory cycle by 9.84\% (absolute) and the best multi-input model by  2.26\% (absolute) in F-score.

\bibliography{refs}
\bibliographystyle{IEEEbib}

\end{document}